\documentclass[11pt]{article}
\usepackage{amssymb}
\author{Keith Anguige}
\title{A class of perfect-fluid cosmologies with polarised Gowdy symmetry and a Kasner-like singularity}
\begin{document}
\maketitle
\begin{abstract}
We prove the existence of a class of perfect-fluid cosmologies which have polarised Gowdy symmetry and a Kasner like singularity. These solutions of the Einstein equations depend on four free functions of one space coordinate and are constructed by solving a system of Fuchsian equations.
\end{abstract}
\section{Introduction}
In this paper we use the Fuchsian algorithm to construct a family of perfect-fluid cosmologies with polarised Gowdy symmetry and with Kasner like asymptotics at early times. This family depends on the maximum number of free functions for spacetimes within the symmetry class. The technique used is to perturb exact Bianchi I solutions in one space-direction and to solve a Fuchsian system of equations for the perturbation. The results obtained are a generalisation of those in \cite{i} in the sense that the symmetry requirement has been relaxed a little, but here we require the free data for the field equations to be analytic rather than merely~$C^{\infty}$.  

The main result is stated as Theorem 6.1 at the end of the paper.

\section{Exact solutions}
The general Bianchi I solution of the Einstein-perfect fluid equations, as given in \cite{ii} is
\begin{equation}ds^{2}=-B^{2(\gamma-1)}d\tau^{2}+\tau^{2p_{1}}B^{2q_{1}}dx^{2}+\tau^{2p_{2}}B^{2q_{2}}dy^{2}+\tau^{2p_{3}}B^{2q_{3}}dz^{2}\end{equation}
where
\begin{equation}B^{2-\gamma}=\alpha+m^{2}\tau^{2-\gamma}~~~~\alpha\geq 0,~m>0\end{equation}
\begin{equation}p_{1}+p_{2}+p_{3}=1~,~p_{1}^{2}+p_{2}^{2}+p_{3}^{2}=1~,~q_{i}=\frac{2}{3}-p_{i}\end{equation}
(The case~$\alpha=0$~is an FRW model and is excluded in what follows.)

The density of the fluid is given by
\begin{equation}\mu=\frac{4m^{2}}{3\tau^{\gamma}B^{\gamma}}\end{equation}
When we come to write down Einstein's equations for a metric with polarised Gowdy symmetry we will use conformal coordinates, i.e coordinates for which the metric takes the form
\begin{equation}ds^{2}=e^{2A(t,x)}(-dt^{2}+dx^{2})+R(t,x)e^{W(t,x)}dy^{2}+R(t,x)e^{-W(t,x)}dz^{2}\end{equation}
The model solution (1) may be written in conformal coordinates  by making the transformation
\begin{equation}\tau\rightarrow t=\int_{0}^{\tau}B^{p_{1}+\gamma-\frac{5}{3}}(s)s^{-p_{1}}~ds\end{equation}
Then the metric (1) takes the form (5) with
\begin{equation}e^{A}=\tau^{p_{1}}B^{\frac{2}{3}-p_{1}}~,~e^{W}=\tau^{p_{2}-p_{3}}B^{p_{3}-p_{2}}~,~R=\tau^{1-p_{1}}B^{(p_{1}+\frac{1}{3})}\end{equation}
and~$\tau$~given implicitly as function of~$t$~ by the relation (6).

\section{The Einstein-perfect fluid equations}
We will assume that spacetime is filled with a polytropic perfect fluid, so that the stress tensor takes the form
\begin{equation}T^{\alpha\beta}=(\rho+P)u^{\alpha}u^{\beta}+Pg^{\alpha\beta}\end{equation}
with~$u^{\alpha}u_{\alpha}=-1$~and~$P=(\gamma-1)\rho,~1<\gamma<2$.

With this stress tensor the Einstein evolution equations for the metric (5) are
\begin{equation}R_{tt}-R_{xx}=Re^{2A}\rho(2-\gamma)\end{equation}
\begin{equation}W_{tt}-W_{xx}+R^{-1}(R_{t}W_{t}-R_{x}W_{x})=0\end{equation}
\begin{equation}A_{tt}-A_{xx}-\frac{1}{4}R^{-2}(R_{t}^{2}-R_{x}^{2})+\frac{1}{4}(W_{t}^{2}-W_{x}^{2})=-\frac{1}{2}\gamma e^{2A}\rho\end{equation}
while the constraints read
\begin{displaymath}R^{-1}R_{xx}-\frac{1}{4}R^{-2}(R_{t}^{2}+R_{x}^{2})-R^{-1}R_{t}A_{t}-R^{-1}R_{x}A_{x}+\frac{1}{4}(W_{t}^{2}+W_{x}^{2})\end{displaymath}
\begin{equation}=-e^{2A}\rho(1+\gamma(v^{1})^{2})\end{equation}
\begin{equation}-R^{-1}R_{tx}+R^{-1}(R_{t}A_{x}+A_{t}R_{x})+\frac{1}{2}R^{-2}R_{t}R_{x}-\frac{1}{2}W_{t}W_{x}=-e^{2A}\gamma\rho v^{0}v^{1}\end{equation}
where~$v^{i}=e^{A}u^{i}$~and hence~$(v^{0})^{2}-(v^{1})^{2}=1$.

The Euler equations~$\nabla^{\alpha}T_{\alpha\beta}=0$~take the following explicit form
\begin{displaymath}\gamma v^{0}(v^{0}\rho_{t}+v^{1}\rho_{x})+(1-\gamma)\rho_{t}\end{displaymath}
\begin{displaymath}+\gamma\rho\{v^{0}(2v^{0}_{t}+A_{x}v^{1})+v^{1}(v^{1}_{x}-v^{1}A_{x}+A_{x}v^{0}+A_{t}v^{1})\}\end{displaymath}
\begin{equation}+\gamma\rho v^{0}\{v^{0}A_{t}+v^{1}_{x}+v^{1}A_{x}+R^{-1}(v^{0}R_{t}+v^{1}R_{x})\}=0\end{equation}
and
\begin{displaymath}\gamma v^{1}(v^{0}\rho_{t}+v^{1}\rho_{x})+(1-\gamma)\rho_{x}+\gamma\rho\{v^{0}(v^{1}_{t}+v^{0}A_{x})\}\end{displaymath}
\begin{equation}+\gamma\rho\{v^{1}(v^{0}_{t}+2v^{1}_{x}+v^{0}(2A_{t}+R^{-1}R_{t})+v^{1}(A_{x}+R^{-1}R_{x})\}=0\end{equation}

\section{Inhomogeneous perturbations}
We now seek solutions ~$(\rho,v^{1},A,R,W)$~of (9)-(15) which depend on the maximum number of free analytic functions, namely four, and which approach the model forms (1)-(4) for each~$x$~as~$t\rightarrow 0$. 

To be specific we make the following ansatz
\begin{equation}A=p_{1}\log{\tau}+\left(\frac{2}{3}-p_{1}\right)\log{B}+t\tilde{A}\end{equation}
\begin{equation}R=\tau^{1-p_{1}}B^{(p_{1}+\frac{1}{3})}(1+\tilde{R})\end{equation}
\begin{equation}W=c(x)+(p_{2}-p_{3})\log{\tau}+(p_{3}-p_{2})\log{B}+t\tilde{W}\end{equation}
\begin{equation}\log{\rho}=\log{\mu}+t^{(\gamma-1-p_{1})/(1-p_{1})+\epsilon}\phi\end{equation}
\begin{equation}v^{1}=\tau^{\gamma-1-p_{1}}(G(x)+t^{\epsilon}\tilde{\psi})\end{equation}
where
\begin{equation}B^{2-\gamma}=\alpha(x)+m^{2}(x)\tau^{2-\gamma}\end{equation}
\begin{equation}G(x)=-\frac{3}{4m^{2}\gamma}\alpha^{(p_{1}+\frac{1}{3})/(2-\gamma)}\times\end{equation}
\begin{equation}\left\{-\frac{1}{2}(p_{2}-p_{3})c_{x}+\frac{\alpha_{x}}{\alpha}\frac{(3\gamma(1-p_{1})+2(3p_{1}-1))}{3(2-\gamma)}+(p_{1})_{x}\left(\frac{2(1-p_{1})}{(2-\gamma)}\log{\alpha}+1\right)\right\}\end{equation}
and~$\tau(t,x)$~is determined by the relation (6).

$\alpha(x)$~and~$m(x)$~are chosen as strictly positive analytic functions,~$c(x)$~is an analytic function and the Kasner exponent~$p_{1}(x)$~is subject to the restriction~$p_{1}(x)<\gamma-1-k$~for some arbitrarily small constant~$k$. The analytic function~$\epsilon(x)$~is chosen to satisfy
\begin{equation}0<\epsilon(x)<\textrm{min}\left(\frac{2-\gamma}{1-p_{1}},\frac{\gamma-1-p_{1}}{1-p_{1}}\right)\end{equation}

We are looking for analytic solutions of the Einstein-perfect fluid equations for which
~$\tilde{A},~\tilde{R},~\tilde{W},~\phi$~and~$\tilde{\psi}$~tend to zero as~$t$~tends to zero. The choice of~$G(x)$~ensures that the momentum constraint is satisfied at~$t=0$.

It is convenient to write the field equations for~$\tilde{A},~\tilde{R},~\tilde{W},~\phi$~and~$\tilde{\psi}$~in first order form. To do this we introduce the following new variables:
\begin{displaymath}U=\tilde{R}_{t},~Q=(t\tilde{A})_{t},~X=\tilde{R}_{x},~Y=t\tilde{A}_{x},~V=(t\tilde{W})_{t},~Z=t\tilde{W}_{x},~S=t^{-1}\tilde{R}\end{displaymath}
In terms of these variables the evolution equations (9)-(11) take the form
\begin{equation}tX_{t}=tU_{x}\end{equation}
\begin{equation}tY_{t}=tQ_{x}\end{equation}
\begin{equation}t\tilde{R}_{t}=tU\end{equation}
\begin{equation}t\tilde{A}_{t}+\tilde{A}-Q=0\end{equation}
\begin{equation}tZ_{t}=tV_{x}\end{equation}
\begin{equation}t\tilde{W}_{t}+\tilde{W}-V=0\end{equation}
\begin{equation}tS_{t}+S-U=0\end{equation}
\begin{displaymath}tU_{t}+2U=tX_{x}+2U\left(1-\left((1-p_{1})B^{2-\gamma}+m^{2}\left(p_{1}+\frac{1}{3}\right)\tau^{2-\gamma}\right)B^{-(p_{1}+\frac{1}{3})}t\tau^{p_{1}-1}\right)\end{displaymath}
\begin{displaymath}+t\tau^{p_{1}-1}B^{-(p_{1}+\frac{1}{3})}\{2X(\tau^{1-p_{1}}B^{p_{1}+\frac{1}{3}})_{x}+(1+\tilde{R})(\tau^{1-p_{1}}B^{p_{1}+\frac{1}{3}})_{xx}\}\end{displaymath}
\begin{equation}+\frac{4m^{2}}{3}(2-\gamma)(1+tS)t\tau^{2p_{1}-\gamma}B^{\frac{4}{3}-2p_{1}-\gamma}(\textrm{exp}(2t\tilde{A}+t^{\epsilon+(\gamma-1-p_{1})(1-p_{1})^{-1}}\phi)-1)\end{equation}

\begin{displaymath}tQ_{t}-\frac{1}{2}U+\frac{1}{2}\frac{(p_{2}-p_{3})}{(1-p_{1})}V=tY_{x}+t\{(p_{1}\log{\tau})_{x}+(\log{B^{\frac{2}{3}-p_{1}}})_{xx}\}\end{displaymath}
\begin{displaymath}+\frac{1}{4}(1+\tilde{R})^{-2}tU^{2}-\frac{1}{2}U\left(1-t\tau^{p_{1}-1}B^{-(p_{1}+\frac{1}{3})}(1+tS)^{-1}\left((1-p_{1})B^{2-\gamma}+m^{2}\left(p_{1}+\frac{1}{3}\right)\tau^{2-\gamma}\right)\right)\end{displaymath}
\begin{displaymath}+\frac{1}{4}t\tau^{2(p_{1}-1)}B^{-2(p_{1}+\frac{1}{3})}\left(\tau^{1-p_{1}}B^{p_{1}+\frac{1}{3}}X+(1+\tilde{R})(\tau^{1-p_{1}}B^{p_{1}+\frac{1}{3}})_{x}\right)^{2}\end{displaymath}
\begin{displaymath}-\frac{1}{4}tV^{2}+\frac{1}{2}V(p_{2}-p_{3})\left((1-p_{1})^{-1}-t\tau^{p_{1}-1}B^{\frac{5}{3}+p_{1}-\gamma}+m^{2}t\tau^{1+p_{1}-\gamma}B^{-\frac{1}{3}-p_{1}}\right)\end{displaymath}
\begin{displaymath}+\frac{1}{4}t\left(c_{x}+((p_{2}-p_{3})\log{\tau})_{x}+((p_{3}-p_{2})\log{B})_{x}+Z\right)^{2}\end{displaymath}
\begin{equation}-\frac{2m^{2}}{3}\gamma B^{\frac{4}{3}-2p_{1}-\gamma}t\tau^{2p_{1}-\gamma}\left(\textrm{exp}(2t\tilde{A}+t^{\epsilon+(\gamma-1-p_{1})(1-p_{1})^{-1}}\phi)-1\right)\end{equation}
\begin{displaymath}tV_{t}+V+\frac{(p_{2}-p_{3})}{(1-p_{1})}U=tZ_{x}+t\{c_{xx}+((p_{2}-p_{3})\log{\tau})_{xx}+((p_{3}-p_{2})\log{B})_{xx}\}\end{displaymath}
\begin{displaymath}+\frac{t\tau^{p_{1}-1}B^{-p_{1}-\frac{1}{3}}}{(1+\tilde{R})}(c_{x}+((p_{2}-p_{3})\log{\tau})_{x}+((p_{3}-p_{2})\log{B})_{x}+Z)(\tau^{1-p_{1}}B^{p_{1}+\frac{1}{3}}X+(1+\tilde{R})(\tau^{1-p_{1}}B^{p_{1}+\frac{1}{3}})_{x})\end{displaymath}
\begin{displaymath}+U\left(\frac{(p_{2}-p_{3})}{(1-p_{1})}-(1+tS)^{-1}((p_{2}-p_{3})t\tau^{p_{1}-1}B^{\frac{5}{3}-p_{1}-\gamma}+m^{2}(p_{3}-p_{2})t\tau^{1+p_{1}-\gamma}B^{-\frac{1}{3}-p_{1}}+tV)\right)\end{displaymath}
\begin{equation}+V(1-B^{-p_{1}-\frac{1}{3}}t\tau^{p_{1}-1}((1-p_{1})B^{2-\gamma}+m^{2}\tau^{2-\gamma}(p_{1}+1/3)))\end{equation}

The Euler equations , after some rearrangement, may be written
\begin{displaymath}\left(\frac{1+2\tau^{2\beta}\psi^{2}}{v^{0}}-\frac{2\gamma v^{0}\tau^{2\beta}\psi^{2}}{1+\gamma\tau^{2\beta}\psi^{2}}\right)(t\tilde{\psi}_{t}+\epsilon\tilde{\psi})=\end{displaymath}
\begin{displaymath}t^{1-\epsilon}\psi\left(\frac{\gamma v^{0}(v^{0}+\tau^{\beta})}{1+\gamma\tau^{2\beta}\psi^{2}}-2\right)(\tau^{\beta}\psi_{x}+\psi(\tau^{\beta})_{x})\end{displaymath}
\begin{displaymath}+t^{1-\epsilon}\tau^{-\beta}\left(\frac{\gamma-1}{\gamma}+\frac{\gamma(v^{0})^{2}\tau^{2\beta}\psi^{2}}{1+\gamma\tau^{2\beta}\psi^{2}}\right)((m^{2})_{x}-\gamma(\tau_{x}+B_{x})+t^{\epsilon+\beta(1-p_{1})^{-1}}(\phi_{x}+\phi(\beta(1-p_{1})^{-1})_{x}\log{t}))\end{displaymath}
\begin{displaymath}+t^{1-\epsilon}v^{0}\psi\left(\frac{\gamma((v^{0})^{2}+\tau^{2\beta}\psi^{2})}{1+\gamma\tau^{2\beta}\psi^{2}}-2\right)((2/3-p_{1})B^{-\frac{1}{3}-p_{1}}m^{2}\tau^{-\beta}+Q)\end{displaymath}
\begin{displaymath}-t^{1-\epsilon}\tau^{-\beta}\left(1+2\tau^{2\beta}\psi^{2}-\frac{\gamma\tau^{\beta}\psi v^{0}}{1+\gamma\tau^{2\beta}\psi^{2}}(3v^{0}\tau^{\beta}\psi-\tau^{2\beta}\psi^{2})\right)((p_{1}\log{\tau})_{x}+(\log{B^{\frac{2}{3}-p_{1}}})_{x}+Y)\end{displaymath}
\begin{displaymath}+t^{1-\epsilon}\psi\left(\frac{\gamma(v^{0})^{3}}{1+\gamma\tau^{2\beta}\psi^{2}}-v^{0}\right)((1+\tilde{R})^{-1}U+m^{2}(p_{1}+1/3)B^{-p_{1}-\frac{1}{3}}\tau^{\beta})\end{displaymath}
\begin{displaymath}+t^{1-\epsilon}\tau^{\gamma-2}\psi^{2}\left(\frac{\gamma(v^{0})^{2}}{1+\gamma\tau^{2\beta}\psi^{2}}-1\right)B^{-p_{1}-\frac{1}{3}}(1+\tilde{R})^{-1}(\tau^{1-p_{1}}B^{p_{1}+\frac{1}{3}}X+(1+\tilde{R})(\tau^{1-p_{1}}B^{p_{1}+\frac{1}{3}})_{x})\end{displaymath}
\begin{displaymath}+t^{1-\epsilon}\tau^{p_{1}-1}\psi B^{\frac{5}{3}-p_{1}-\gamma}\left\{\beta\left(1+\frac{2\gamma v^{0}\tau^{2\beta}\psi^{2}}{1+\gamma\tau^{2\beta}\psi^{2}}-\frac{1+2\tau^{2\beta}\psi^{2}}{v^{0}}\right)\right.\end{displaymath}
\begin{equation}\left.+p_{1}\left(\gamma\left(\frac{v^{0}((v^{0})^{2}+\tau^{2\beta}\psi^{2})}{1+\gamma\tau^{2\beta}\psi^{2}}-1\right)+2(1-v^{0})\right)+(1-p_{1})\left(\gamma\left(\frac{(v^{0})^{2}}{1+\gamma\tau^{2\beta}\psi^{2}}-1\right)+1-v^{0}\right)\right\}\end{equation}
and
\begin{displaymath}\left(1+\gamma\tau^{2\beta}\psi^{2}-\frac{2\gamma(v^{0})^{2}\tau^{2\beta}\psi^{2}}{1+2\tau^{2\beta}\psi^{2}}\right)t\phi_{t}+\left(\epsilon+\frac{\beta}{1-p_{1}}\right)\phi=\end{displaymath}
\begin{displaymath}=\gamma\left\{(\epsilon+\beta(1-p_{1})^{-1})\phi\left(\frac{1}{\gamma}+\frac{2(v^{0})^{2}\tau^{2\beta}\psi^{2}}{1+2\tau^{2\beta}\psi^{2}}-\frac{1+\gamma\tau^{2\beta}\psi^{2}}{\gamma}\right)\right.\end{displaymath}
\begin{displaymath}+\gamma t^{1+\frac{\beta}{p_{1}-1}-\epsilon}\tau^{p_{1}-1}\tau^{2\beta}\psi^{2}\left(1-\frac{2(v^{0})^{2}}{1+2\tau^{2\beta}\psi^{2}}\right)(B^{\frac{5}{3}-p_{1}-\gamma}+m^{2}\tau^{2-\gamma}B^{-\frac{1}{3}-p_{1}})\end{displaymath}
\begin{displaymath}-(t\phi_{x}+t\phi\log{t}(\epsilon+\beta(1-p_{1})^{-1})_{x}+t^{1+\frac{\beta}{p_{1}-1}-\epsilon}((m^{2})_{x}-\gamma(\tau_{x}+B_{x})))v^{0}\tau^{\beta}\psi\end{displaymath}
\begin{displaymath}\times\left(1+\frac{2}{1+2\tau^{2\beta}\psi^{2}}\left(\frac{\gamma-1}{\gamma}-\tau^{2\beta}\psi^{2}\right)\right)\end{displaymath}
\begin{displaymath}+t^{1+\frac{\beta}{p_{1}-1}-\epsilon}\left(\frac{4\tau^{2\beta}\psi^{2}v^{0}}{1+2\tau^{2\beta}\psi^{2}}-v^{0}-\tau^{\beta}\psi\right)((\tau^{\beta}G)_{x}+\tau^{\beta}t^{\epsilon}\tilde{\psi}_{x}+(\tau^{\beta}t^{\epsilon})_{x}\tilde{\psi})\end{displaymath}
\begin{displaymath}+t^{1+\frac{\beta}{p_{1}-1}-\epsilon}\left(\frac{4(v^{0})^{2}\tau^{2\beta}\psi^{2}}{1+2\tau^{2\beta}\psi^{2}}-1-2\tau^{2\beta}\psi^{2}\right)Q\end{displaymath}
\begin{displaymath}+t^{1+\frac{\beta}{p_{1}-1}-\epsilon}\tau^{2\beta}\psi^{2}\left(\frac{4(v^{0})^{2}}{1+2\tau^{2\beta}\psi^{2}}-2\right)(p_{1}\tau^{p_{1}-1}B^{\frac{5}{3}-p_{1}-\gamma}+((2/3)-p_{1})B^{-\frac{1}{3}-p_{1}}m^{2}\tau^{-\beta})\end{displaymath}
\begin{displaymath}+t^{1+\frac{\beta}{p_{1}-1}-\epsilon}\left(\tau^{\beta}\psi(\tau^{\beta}\psi-v^{0})((p_{1}\log{\tau})_{x}+(\log{B^{\frac{2}{3}-p_{1}}})_{x}+Y)+\frac{(v^{0}\tau^{\beta}\psi-(v^{0})^{2})}{1+\tilde{R}}U\right)\end{displaymath}
\begin{displaymath}+t^{1+\frac{\beta}{p_{1}-1}-\epsilon}\tau^{2\beta}\psi^{2}\left(\frac{2(v^{0})^{2}}{1+2\tau^{2\beta}\psi^{2}}-1\right)\tau^{p_{1}-1}B^{-p_{1}-\frac{1}{3}}((1-p_{1})B^{2-\gamma}+m^{2}(p_{1}+1/3)\tau^{2-\gamma})\end{displaymath}
\begin{equation}+\left.t^{1+\frac{\beta}{p_{1}-1}-\epsilon}\tau^{\gamma-2}\psi\left(\frac{2v^{0}\tau^{2\beta}\psi^{2}}{1+2\tau^{2\beta}\psi^{2}}-v^{0}\right)\frac{B^{-p_{1}-\frac{1}{3}}}{1+\tilde{R}}(\tau^{1-p_{1}}B^{p_{1}+\frac{1}{3}}X+(1+\tilde{R})(\tau^{1-p_{1}}B^{p_{1}+\frac{1}{3}})_{x})\right\}\end{equation}

\section{Existence and uniqueness of solutions}
A careful inspection of the field equations (25)-(36) shows that they may be written in the form 
\begin{equation}t\partial_{t}u+N(x)u=t^{\delta}H(t,x,u,u_{x})\end{equation}
where~$u$~stands for~$(U,\tilde{A},Q,X,Y,\tilde{R},V,Z,S,\phi,\tilde{\psi})$,~$\delta$~is a strictly positive constant and~$H$~is continuous in~$t$~and analytic in its other arguments. The matrix~$N(x)$~has positive eigenvalues.

It follows by \cite{iii} that (37) has a unique analytic solution~$u$~with~$u(0)=0$

\section{The constraints}
Define constraint quantities~$C_{1},~C_{0}$~by
\begin{displaymath}C_{0}=R^{-1}R_{xx}-\frac{1}{4}R^{-2}(R_{t}^{2}+R_{x}^{2})-R^{-1}R_{t}A_{t}-R^{-1}R_{x}A_{x}+\frac{1}{4}(W_{t}^{2}+W_{x}^{2})\end{displaymath}
\begin{equation}+e^{2A}\rho(1+\gamma(v^{1})^{2})\end{equation}
\begin{equation}C_{1}=-R^{-1}R_{tx}+R^{-1}(R_{t}A_{x}+A_{t}R_{x})+\frac{1}{2}R^{-2}R_{t}R_{x}-\frac{1}{2}W_{t}W_{x}+e^{2A}\gamma\rho v^{0}v^{1}\end{equation}
If the evolution equations (25)-(36) are satisfied then a calculation shows that the following hold
\begin{equation}\partial_{t}C_{0}=-\partial_{x}C_{1}-\frac{R_{t}}{R}C_{0}-\frac{R_{x}}{R}C_{1}\end{equation}
\begin{equation}\partial_{t}C_{1}=-\partial_{x}C_{0}-\frac{R_{t}}{R}C_{1}-\frac{R_{t}}{R}C_{0}\end{equation}
One also calculates that the quantities~$\tilde{C}_{0}=tC_{0},~\tilde{C}_{1}=tC_{1}$~tend to zero as~$t$~tends to zero. These quantities satisfy the following\begin{equation}t\partial_{t}\tilde{C}_{0}+\left(\frac{tR_{t}}{R}-1\right)\tilde{C}_{0}=-t\partial_{x}\tilde{C}_{1}-\frac{tR_{x}}{R}\tilde{C}_{1}\end{equation} 
\begin{equation}t\partial_{t}\tilde{C}_{1}+\left(\frac{tR_{t}}{R}-1\right)\tilde{C}_{1}=-t\partial_{x}\tilde{C}_{0}-\frac{tR_{x}}{R}\tilde{C}_{0}\end{equation}
Now~$R_{x}/R$~is~$O(1)$~and~$tR_{t}/R=1+O(t^{\delta})$~for some~$\delta>0$. It thus follows that~$\tilde{C}_{0}$~and~$\tilde{C}_{1}$~are identically zero and thus the constraints are satisfied.

Summarising the results of sections (4)-(6) we have  proved the following

\textbf{Theorem 6.1} Given two strictly positive analytic functions~$\alpha(x),~m(x)$, an analytic function~$c(x)$~and an analytic function~$p_{1}(x)$~satisfying
\begin{equation}-\frac{1}{3}\leq p_{1}(x)<\gamma-1-k\end{equation}
for some small~$k>0$~, there exists a unique solution~$(g_{\mu\nu},\rho,u^{\alpha})$~of the Einstein equations coupled to a~$\gamma$-law perfect fluid on~$\mathbb{R}^{3}\times (0,T)$~satisfying
\begin{displaymath}ds^{2}=e^{2A}(t,x)(-dt^{2}+dx^{2})+R(t,x)e^{W(t,x)}(dy)^{2}+R(t,x)e^{-W(t,x)}(dz)^{2}\end{displaymath}
\begin{displaymath}A=p_{1}\log{t}+\log{B^{\frac{2}{3}-p_{1}}}+O(t)\end{displaymath}
\begin{displaymath}R=\tau^{1-p_{1}}B^{p_{1}+\frac{1}{3}}(1+O(t))\end{displaymath}
\begin{displaymath}W=c(x)+(p_{2}-p_{3})\log{t}+(p_{3}-p_{2})\log{B}+O(t)\end{displaymath}
\begin{displaymath}\log{\rho}=\log{\frac{4m^{2}}{3\tau^{\gamma}B^{\gamma}}}+O(t^{(\gamma-1-p_{1})(1-p_{1})^{-1}})\end{displaymath}
\begin{displaymath}v^{1}=O(t^{(\gamma-1-p_{1})(1-p_{1})^{-1}})\end{displaymath}
where~$\tau$~is implicitly given by
\begin{displaymath}t=\int_{0}^{\tau}B^{p_{1}+\gamma-\frac{5}{3}}(s)s^{-p_{1}}~ds.\end{displaymath}

\end{document}